\makeatletter
\newcommand{\dontusepackage}[2][]{%
  \@namedef{ver@#2.sty}{9999/12/31}%
  \@namedef{opt@#2.sty}{#1}}
\makeatother
\dontusepackage{subfigure}

\documentclass[]{sigchi}

 \toappear{\copyright~2016 Association for Computing Machinery. ACM acknowledges that this contribution was authored or co-authored by an employee, contractor or affiliate of a national government. As such, the Government retains a nonexclusive, royalty-free right to publish or reproduce this article, or to allow others to do so, for Government purposes only. \\
 {\emph{TEI '16}}, February 14-17, 2016, Eindhoven, Netherlands \\
 Copyright \copyright~2016 ACM. ISBN 978-1-4503-3582-9/16/02 ...\$15.00 \\
 DOI: http://dx.doi.org/10.1145/2839462.2839486}

\usepackage{lmodern}
\usepackage{amssymb,amsmath}
\usepackage{ifxetex,ifluatex}
\usepackage[usenames,dvipsnames]{color}
\usepackage{fixltx2e} 
\ifnum 0\ifxetex 1\fi\ifluatex 1\fi=0 
  \usepackage[utf8]{inputenc}
\else 
  \ifxetex
    \usepackage{mathspec}
    \usepackage{xltxtra,xunicode}
  \else
    \usepackage{fontspec}
  \fi
  \defaultfontfeatures{Mapping=tex-text,Scale=MatchLowercase}
  
\fi
\IfFileExists{upquote.sty}{\usepackage{upquote}}{}
\IfFileExists{microtype.sty}{%
\usepackage{microtype}
\UseMicrotypeSet[protrusion]{basicmath} 
}{}
\usepackage{listings}
\usepackage{graphicx}
\makeatletter
\def\maxwidth{\ifdim\Gin@nat@width>\linewidth\linewidth\else\Gin@nat@width\fi}
\def\maxheight{\ifdim\Gin@nat@height>\textheight\textheight\else\Gin@nat@height\fi}
\makeatother
\setkeys{Gin}{width=\maxwidth,height=\maxheight,keepaspectratio}
\usepackage[font={small,bf},skip=2pt]{caption}
\usepackage{float}
\def\plainauthor{Renaud Gervais, Jérémy Frey, Alexis Gay, Fabien Lotte and Martin Hachet}

\ifxetex
  \usepackage[setpagesize=false, 
              unicode=false, 
              xetex]{hyperref}
\else
  \usepackage[unicode=true]{hyperref}
\fi
\hypersetup{breaklinks=true,
            bookmarks=true,
            pdfauthor={\plainauthor},
            pdftitle={TOBE: Tangible Out-of-Body Experience},
            colorlinks=true,
            citecolor=black,
            urlcolor=blue,
            linkcolor=black,
            pdfborder={0 0 0}}
\usepackage{balance}

\title{TOBE: Tangible Out-of-Body Experience}

\numberofauthors{5}
\author{
  \alignauthor Renaud Gervais\thanks{Co-first authorship, both authors contributed equally to this work.}\\
    \affaddr{Inria, France}\\
    \email{renaud.gervais@inria.fr}\\ 
  \alignauthor Jérémy Frey$^{*}$\\
    \affaddr{Univ. Bordeaux, France}\\
    \email{jeremy.frey@inria.fr}\\
  \alignauthor Alexis Gay\\
    \affaddr{Univ. Bordeaux \mbox{Montaigne, France}}\\ 
    \email{alexis.gay@inria.fr}\\
  \alignauthor Fabien Lotte\\
    \affaddr{Inria, France}\\
    \email{fabien.lotte@inria.fr}\\
  \alignauthor Martin Hachet\\
    \affaddr{Inria, France}\\
    \email{martin.hachet@inria.fr}\\
}

\date{}

\usepackage{times}

\usepackage{suffix}

\usepackage{subfig}

\begin{document}


\maketitle

\begin{abstract}
We propose a toolkit for creating Tangible Out-of-Body Experiences:
exposing the inner states of users using physiological signals such as
heart rate or brain activity. Tobe can take the form of a tangible
avatar displaying live physiological readings to reflect on ourselves
and others. Such a toolkit could be used by researchers and designers to
create a multitude of potential tangible applications, including (but
not limited to) educational tools about Science Technologies Engineering
and Mathematics (STEM) and cognitive science, medical applications or
entertainment and social experiences with one or several users or Tobes
involved. Through a co-design approach, we investigated how everyday
people picture their physiology and we validated the acceptability of
Tobe in a scientific museum. We also give a practical example where two
users relax together, with insights on how Tobe helped them to
synchronize their signals and share a moment.
\end{abstract}

\keywords{
      Physiological Computing;
      Tangible Interaction;
      Spatial Augmented Reality;
      EEG;
      ECG;
      EDA}

      \category{H.5.1}{Multimedia Information Systems}{Artificial, augmented, and virtual realities}
      \category{H.5.2}{User Interfaces}{Prototyping}

\def \citep {\protect\cite}



\makeatletter
\def\url@leostyle{%
  \@ifundefined{selectfont}{\def\UrlFont{\sf}}{\def\UrlFont{\small\bf\ttfamily}}}
\makeatother
\urlstyle{leo}

\def\pprw{8.5in}
\def\pprh{11in}

\setlength{\paperwidth}{\pprw}
\setlength{\paperheight}{\pprh}
\setlength{\pdfpagewidth}{\pprw}
\setlength{\pdfpageheight}{\pprh}

\definecolor{linkColor}{RGB}{6,125,233}
\hypersetup{%
  bookmarksnumbered,
  colorlinks,
  citecolor=black,
  filecolor=black,
  linkcolor=black,
  urlcolor=linkColor,
  breaklinks=true,
}

\WithSuffix\newcommand\caption*{\caption}

\newcommand{\leveldown}
  {\let\section\subsection%
   \let\subsection\subsubsection%
   \let\subsubsection\paragraph%
   \let\paragraph\subparagraph%
  }
  
\newcommand{\levelup}
  {\let\subparagraph\paragraph%
   \let\paragraph\subsubsection%
   \let\subsubsection\subsection%
   \let\subsection\section%
  }

\section{Introduction}\label{introduction}

\begin{figure}
\centering
\includegraphics[]{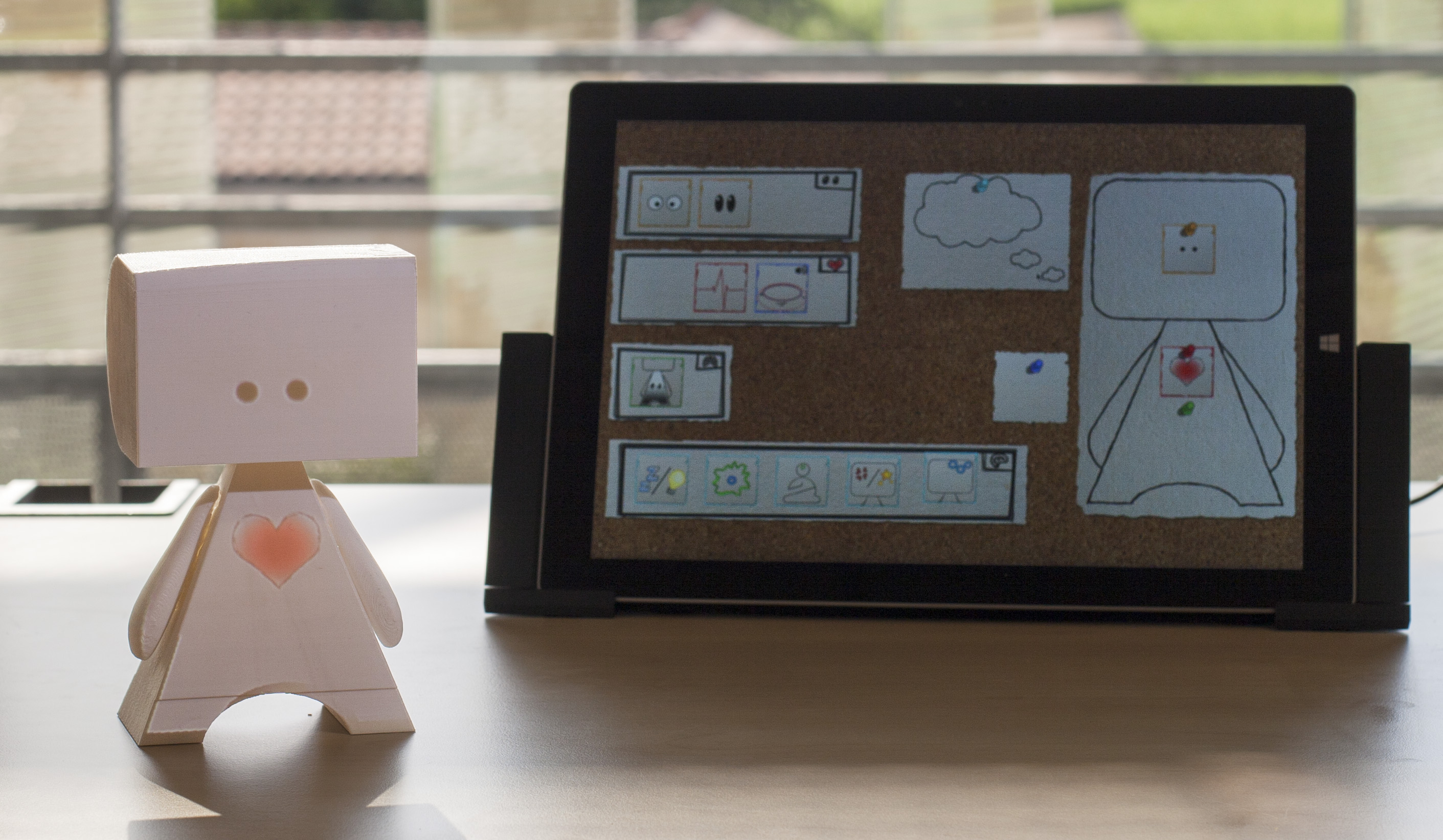}
\caption*{Tobe, the tangible avatar displaying real-time physiological
readings along with the interface to control the different
visualizations.}\label{fig:trailer}
\end{figure}

Wearable computational devices are more accessible and more popular than
ever. These devices are personal and are embedded with physiological
sensors, i.e.~sensors that can monitor signals such as heart beats.
Nowadays even brain activity is within reach of consumers thanks to
cheap alternatives to medical equipment, such as the Emotiv
Epoc\footnote{\url{https://emotiv.com/}} or, closer to the
Do-It-Yourself community, the OpenBCI board\footnote{\url{http://www.openbci.com/}}.
Physiological computing is becoming mainstream. However, for the general
public, the use of such sensors seems mostly centered around
performance. Despite an era of personal development, well-being and
communication, how many smart watches and heart rate belts advertise
themselves as sportspersons' best buddies, while they can account for so
much more than \emph{physical} health? Indeed, physiological computing
is mature enough to asses mental states
\citep{Fairclough2009a, Picard1995, Zander2011, Frey2014a}. Therefore,
it could be used as a mean to better know our own self and others.

On the one hand, physiological technologies are not exploited to their
full potential, on the other hand, we have end users that ignore what
technology has to offer for their well-being. Some companies are
pioneers, as for example Empatica and its Embrace smart watch\footnote{\url{https://www.empatica.com/}},
but they focus on health applications and, consequently, the targeted
consumers are still a niche. Both a process that will raise public
awareness and a collection of meaningful use cases are missing. Finally,
when bodily activity and mental states are at stake -- which are
difficult to conceptualize and often difficult to perceive -- the
feedback given to users matters for them to comprehend at first sight
what is being measured. How to represent the arousal state of someone?
How would \emph{you} represent cognitive workload? We found little
examples besides pies and charts, which are not always obvious
informants in data visualizations -- e.g. \citep{McCandless2010}.

To address these issues, we first conducted surveys and interviews to
gain insight about physiological feedback. We then created Tobe, a
Tangible Out-of-Body Experience shaped as a tangible avatar (Figure
\ref{fig:trailer}). This avatar lets users freely explore and represent
their physiological signals, displayed on the avatar itself using
spatial augmented reality. The overarching goal is to help one reflect
on her physiological and mental states in \emph{her} own way. The main
activity would be for users to actively \emph{build} from the ground up
their own self-representation and then visualize physiological signals
through it. As such, we designed a modular toolkit around Tobe that can
be used to customize any part of the system. Tobe has been tested on two
different occasions in a scientific museum to collect user feedback. A
specialized version of the system was also built to give biofeedback to
multiple users in a relaxation task. Beside these two implementations,
we identified potential future uses of the system, such as a biofeedback
device for stroke rehabilitation or replaying inner states synchronized
along with videos of cherished memories. The latter example could help
create more cherishable versions of personal digital data
\citep{Golsteijn2012}.

Previous works do not embrace such system as a whole and are limited
either to low-level signals or to emotions. Wearables were used in
\citep{Williams2015} to mediate affect using multimodal stimuli (sounds,
heat, vibration, \ldots{}). However, as with the ``Social Skin'' project
\citep{Ugur2013} -- that also embodies emotions into actuated wearables
--, the information given to those around was rather implicit. When a
more comprehensible feedback was studied, as in \citep{Norooz2015}, it
was limited to anatomical models, for instance to teach children how the
body works. Tobe, on the other hand, gives both access to meaningful
visualizations and to additional cognitive states. Tangible proxies and
material representations were already studied in \citep{Khot2014},
although the feedback was not dynamic and, once again, constrained to
bodily activity. While we previously used a tangible puppet as a proxy
for brain activity \citep{Frey2014b}, the settings concerned scientific
outreach and the feedback focused only on preprocessed brain signals,
not on higher level mental states. Our toolkit pushes further the
boundaries of the applications by giving access to physiological
signals, high-level mental states as well as dynamic and customizable
feedback.

Our contribution for this paper consists of a toolkit enabling users to
create an animated tangible representation of their inner states. The
toolkit encompasses the whole workflow, including the physical avatar
creation, sensors, signal processing, feedback and augmentation. It was
tested through two use cases, in public settings and with a multiple
users scenario.

\section{Representing Physiological
Signals}\label{representing-physiological-signals}

Exposing physiological signals in a way that makes sense for the user is
not trivial. Some types of signals might be more obvious to represent
than others. For example, heart activity could be understood using a
symbolic heart shape due to largely accepted cultural references. This
question is, however, harder when talking about more abstract mental
states such as workload. Nevertheless, even the dynamic representation
of low-level physiological signals is still an open question at the
moment \citep{Chanel2015}. We conducted two surveys to gain more insight
about the knowledge and representations people had of different types of
signals and high-level mental states.

In the first survey, conducted online, we asked 36 persons about their
knowledge of physiological signals in general. We inquired about the
self-awareness of inner states on a 7-points Likert scale (1: no
awareness, 7: perfectly aware). About ``internal physiological
activities'', the average score was 3.5 (SD=1.4) and for ``mental
states'', the average score was 4.9 (SD=1.3). The latter score indicates
that the participants were confident about their inner states, even
though a whole literature demonstrates how difficult this is
\citep{Nisbett1977}. Interestingly, we also observed that most of the
participants reduced mental states and physiology to emotions only.
Mentions of any cognitive processes such as vigilance and workload were
very rare (7 out of 36). This lack of knowledge about the inner self and
the different cognitive processes is an opportunity to raise awareness
of the general public about the complexity of the mind. When inquired
about possible uses of a Tobe system, few respondents (6 out of 36) gave
examples other than sports or health. This emphasizes the fact that the
general public is unaware of possibilities of technology for well-being.

\begin{figure}
\centering
\includegraphics[]{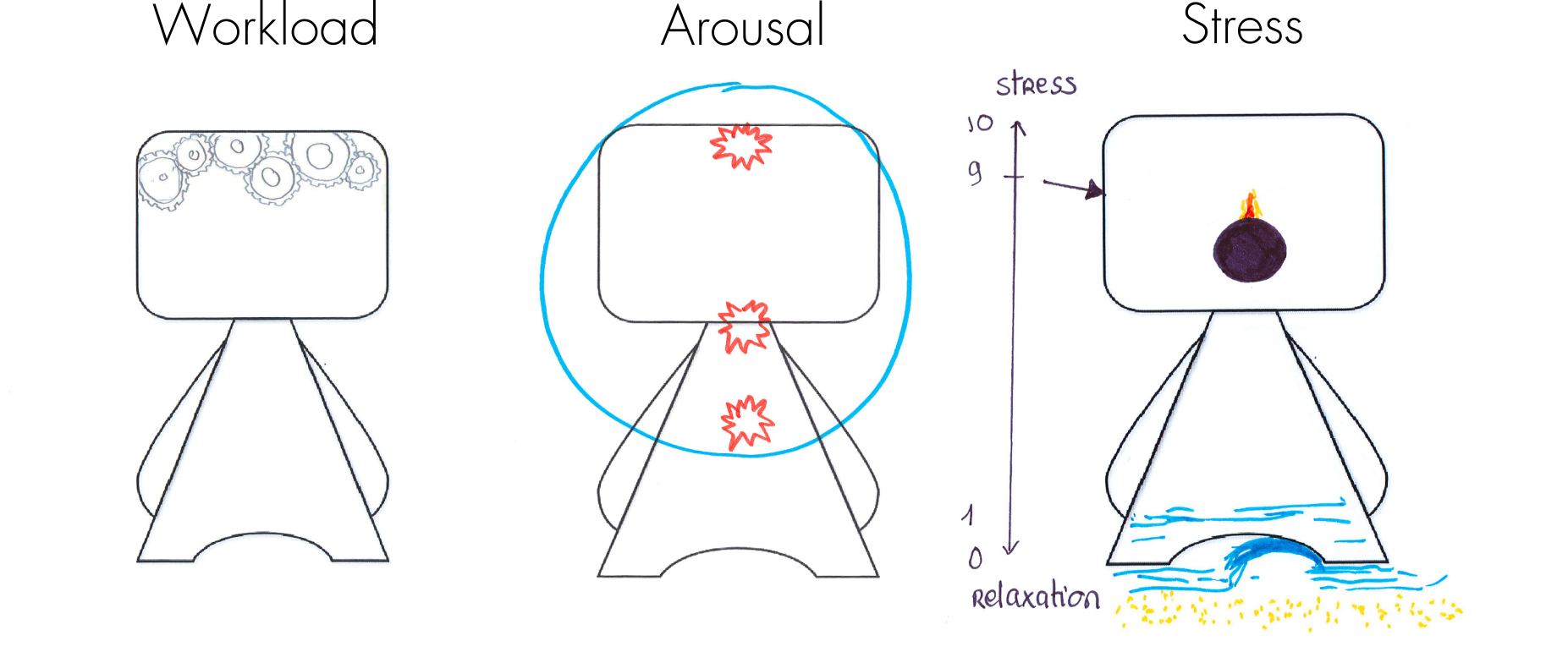}
\caption*{Sample of the drawings made by participants to represent
various high-level metrics.}\label{fig:drawings}
\end{figure}

The second survey specifically investigated how users would shape the
feedback. We focused on visual cues because it was easier to express on
paper, but note that other modalities could be explored, such as sound
\citep{Mealla2011}. We asked 15 participants to express with drawings
and text how they would represent various metrics (Figure
\ref{fig:drawings}). There was little resemblance between subjects for a
given high-level metric and even low-level ones -- breathing and heart
activity -- sprang different views. For example, some people drew a
physiologically accurate heart instead of a simple sketch. Overall,
there was a wide variety of sketches and people were very creative. This
highlighted the absence of consensus on how we conceive and view our
inner states. Therefore, people could benefit from being able to tailor
a meaningful and personal feedback.

\section{Toolkit}\label{toolkit}

We created a tangible anthropomorphic avatar, named Tobe as a host for
displaying real-time feedback. We chose this form factor because we
found evidence in the literature that this combination of
anthropomorphism and tangibility can foster social presence and
likability \citep{Schmitz2010, Hornecker2011}. This also reminds users
and observers that the feedback is linked to an actual being; it helps
to recognize Tobe as a persona and to bond with it, hence it facilitates
engagement.

Our implementation uses open or low-cost hardware and we are releasing
as open-source software the entire pipeline\footnote{\url{https://github.com/potioc/tobe}},
thus facilitating reproduction and dissemination.

\subsection{General Approach}\label{general-approach}

\begin{figure}
\centering
\includegraphics[]{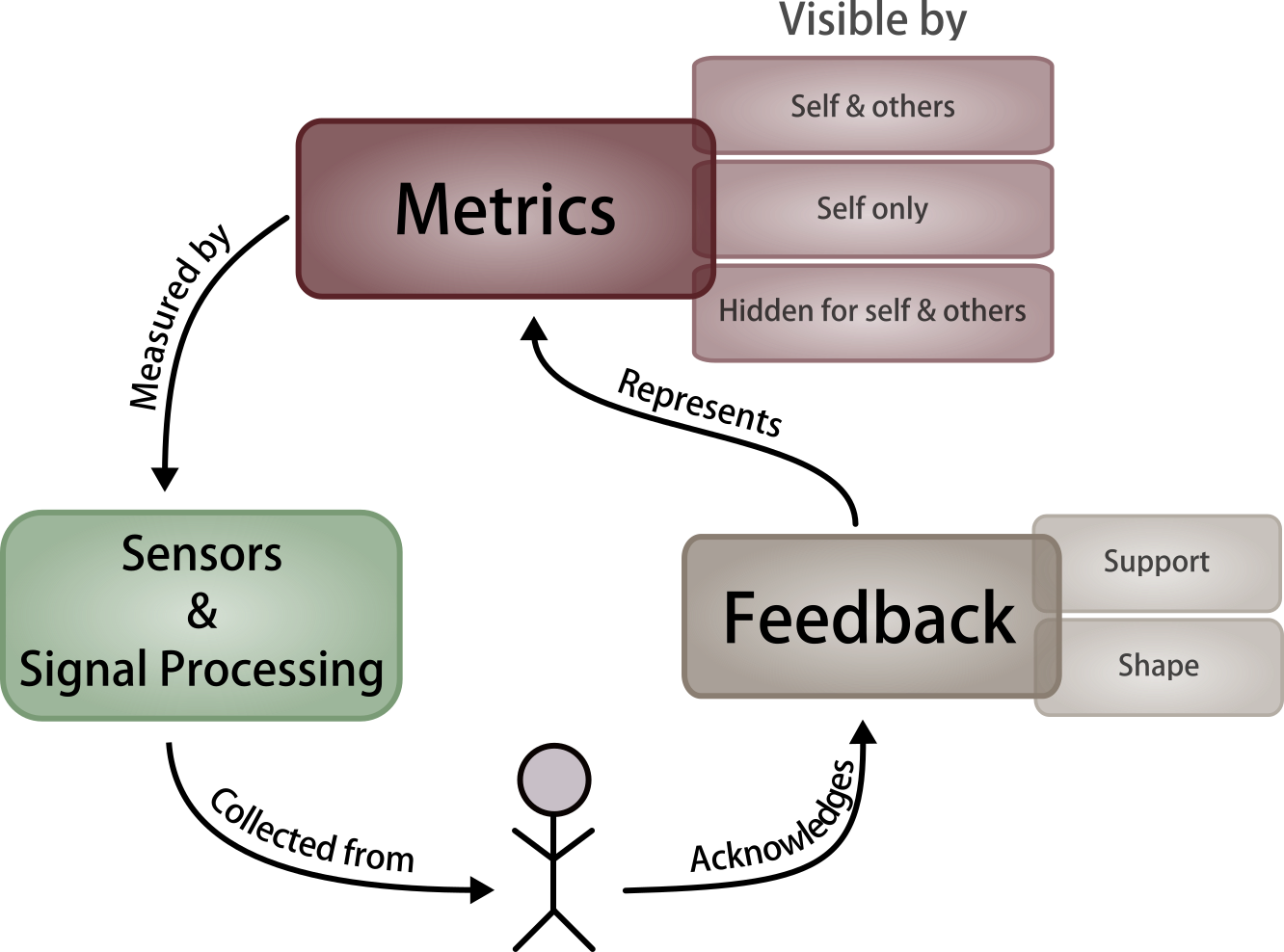}
\caption*{Simplified view of the toolkit that supports
Tobe.}\label{fig:toolkit}
\end{figure}

We conceived a toolkit to assist the creation of representations of
inner activities -- our body at large and the hidden processes of our
mind in particular, making it visible to oneself and to others. The
different components are highlighted in Figure \ref{fig:toolkit}. The
first step consists in choosing a \emph{metric}, e.g.~the arousal level.
For this given metric there are different ways to measure it, that
include a combination of one or multiple \emph{sensor(s) and signal
processing} algorithm(s). One chooses a support to express this metric
(e.g.~tangible avatar, screen, speaker for sound) and creates a shape
associated to it (e.g.~a circle with a changing color, a rhythmic tone).
The conjunction of both the shape and the support produces the
\emph{feedback}. It is an iterative process because when one
acknowledges the feedback, it changes one's self-representation.
Moreover, it creates a feedback loop which affects one's biosignals.

In order to help users mold the system to their likening, we identified
three different degrees of freedom:

\begin{itemize}
\itemsep1pt\parskip0pt\parsep0pt
\item
  The measured physiological signal or mental state (Metric)
\item
  The form factor (Support)
\item
  The display of the signals (Shape)
\end{itemize}

\subsection{Metrics}\label{metrics}

There is a continuum in the visibility of the signals and mental states
measured from physiological sensors, i.e.~metrics. We categorized those
metrics in three different levels, depending on who can perceive them
without technological help.

\begin{enumerate}
\def\labelenumi{\arabic{enumi}.}
\item
  Perceived by self and others, e.g.~eye blinks. Even if those signals
  may sometimes appear redundant as one may directly look at the person
  in order to see them, they are crucial in associating a feedback to a
  user.
\item
  Perceived only by self, e.g.~heart rate or breathing. Mirroring these
  signals provides presence towards the feedback.
\item
  Hidden to both self and others, e.g.~mental states such as cognitive
  workload. This type of metrics holds the most promising applications
  since they are mostly unexplored.
\end{enumerate}

Lower levels (1 \& 2) help to breath life into a proxy used to mediate
the inner state of the user. These metrics are accessible to our
conscious selves. On the other hand, level 3 metrics are little known
and are hard to conceptualize for the general public \citep{Nisbett1977}
and would benefit the most of a system enabling their visualization.

\subsection{Sensors and Signal
Processing}\label{sensors-and-signal-processing}

\begin{figure*}
    \centering
    \subfloat[\label{sealB}]{\includegraphics[width=0.24\hsize]{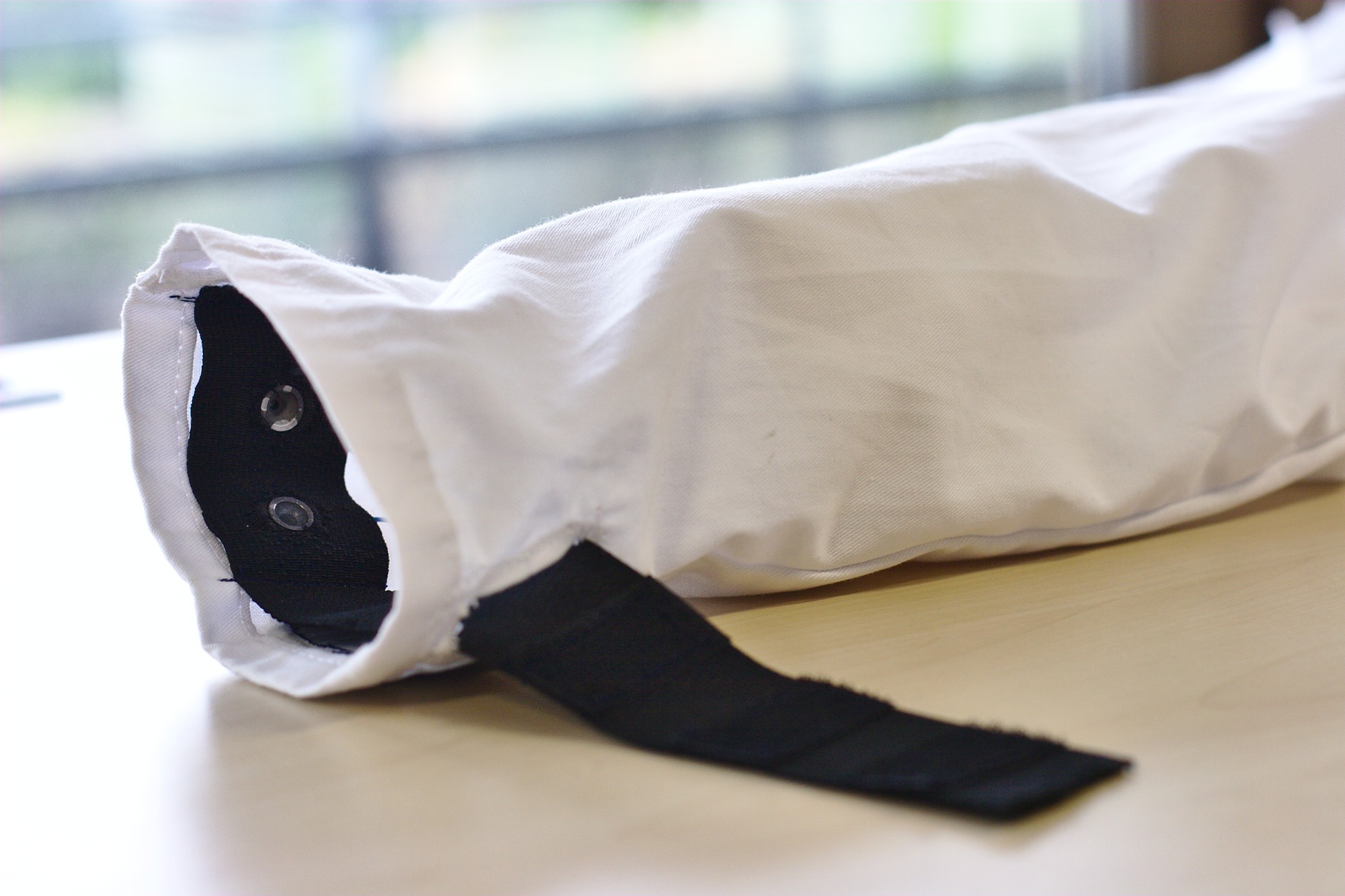}\label{fig:ecg}}\hfill
    \subfloat[\label{sealA}]{\includegraphics[width=0.24\hsize]{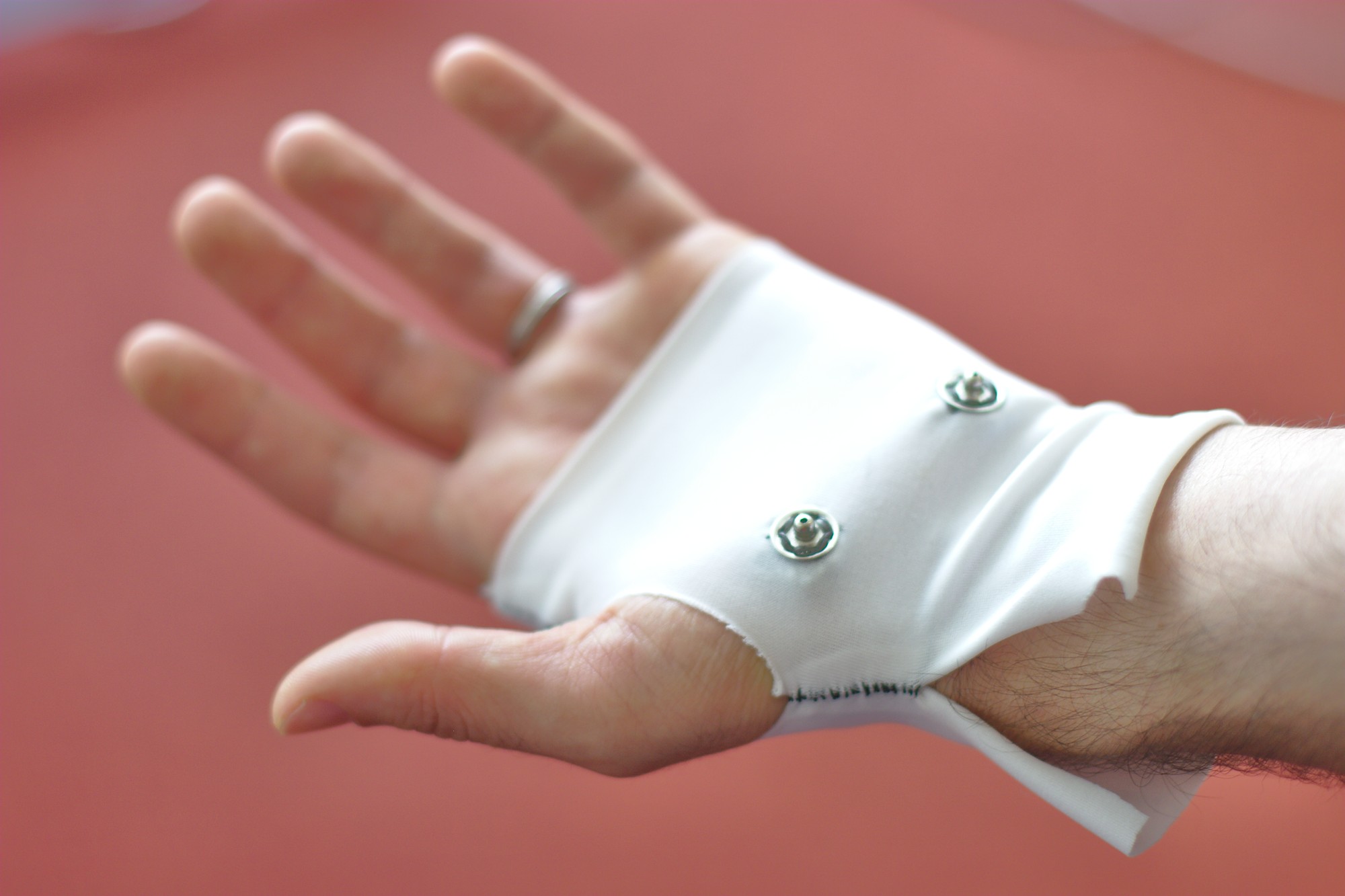}\label{fig:eda}}\hfill
    \subfloat[\label{sealB}]{\includegraphics[width=0.24\hsize]{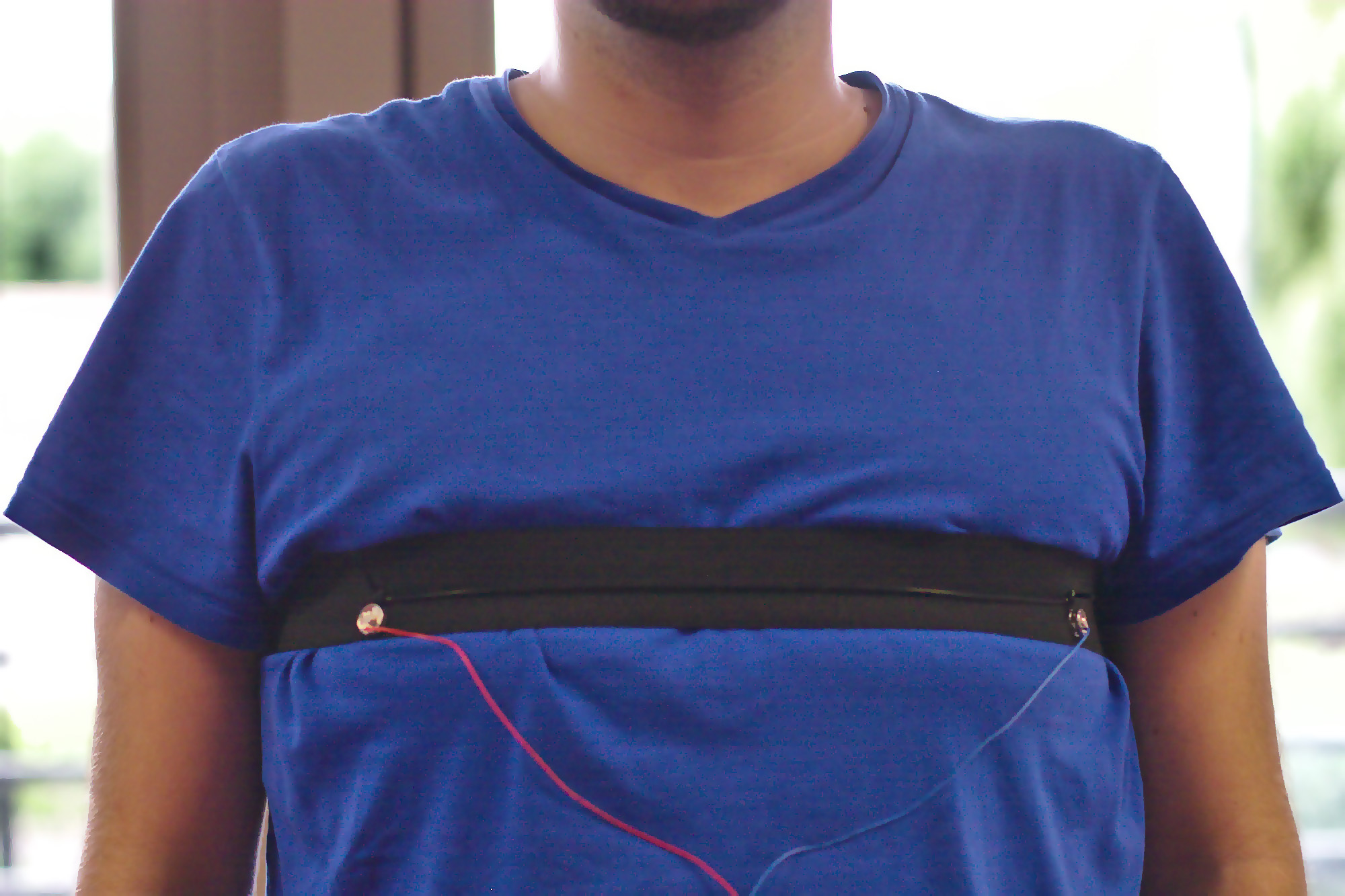}\label{fig:breathing}}\hfill
    \subfloat[\label{sealB}]{\includegraphics[width=0.24\hsize]{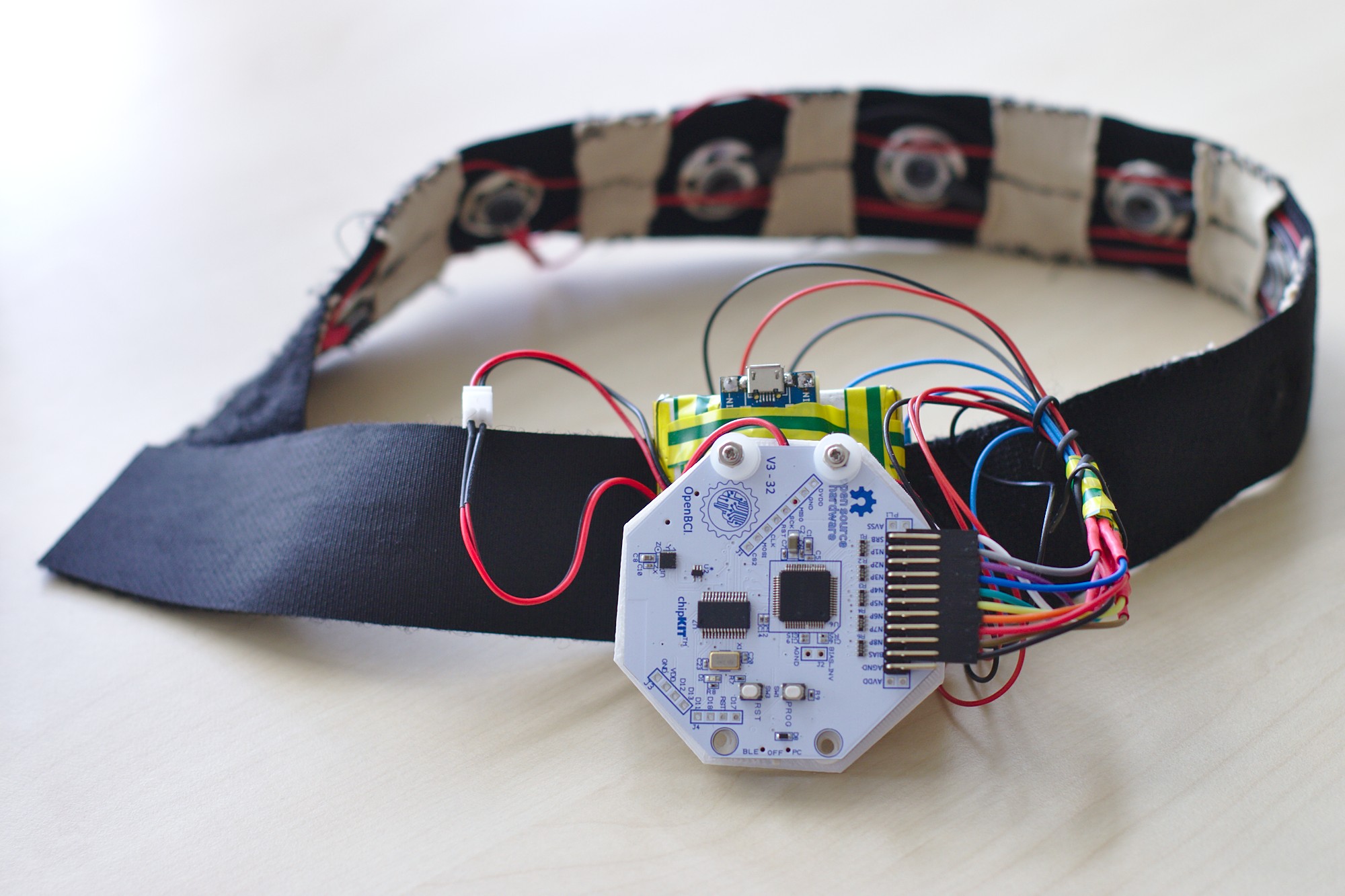}\label{fig:openbci}}
    \caption{Wearables. \emph{a}: coat embedding ECG sensors; \emph{b}: fingerless glove measuring EDA; \emph{c}: breathing belt; \emph{d}: EEG headband.}\label{fig:wearables}
\end{figure*}

Metrics were acquired from five physiological signals. We measured
thoracic circumference for breathing, electrocardiography (ECG) for
heart rate, electrodermal activity (EDA, i.e.~perspiration) for arousal,
electrooculography (EOG, eyes activity) for eye blinks, and
electroencephalography (EEG, brain activity) for most high-level mental
states -- i.e.~vigilance, workload, meditation and valence.

We created the sensors with a wearable form factor in mind. Since we
used Tobe in public settings, it was important that the sensors were
non-invasive (no need to remove clothes or apply gel to the skin) and be
quick to install and remove, while being able to acquire a reliable
signal. With the setup described in this section, we were able to equip
the users and record physiological signals in less than two minutes.

The different sensors were embedded inside a lab coat (Figure
\ref{fig:capsciences}) which could be put on quickly over daily clothes.
This form factor provides enough room in the sleeves and the pockets to
take care of the wiring and electronic components storage. The recording
of the low-level physiological signals (i.e.~everything except EEG) is
done using the BITalino board, an Arduino-based recording device. It
contains modules that amplify various physiological signals and it
embeds a Bluetooth adapter as well as a battery to work in ambulatory
settings. Each physiological signal or mental state index was sent to
the other stage of the toolkit using LSL\footnote{\url{https://github.com/sccn/labstreaminglayer/}},
a network protocol dedicated to physiological recordings.

\subsubsection{ECG}\label{ecg}

We chose to use ECG for heart rate activity as it is more accurate than
light emission-based methods to detect individual heartbeats
\citep{Kranjec2014}. Existing solutions for ECG require electrodes to be
put directly on the chest, e.g.~heart rate monitor belts. We instead
opted for installing TDE-201 Ag-AgCl electrodes from Florida Research
Instruments on both wrists of the user (ECG needs two electrodes
diametrically opposed to sense heart electrical activity). The
electrodes were attached to an elastic band sewn inside the end of the
lab coat sleeves which could be tightened with velcro straps (Figure
\ref{fig:ecg}). ECG was recorded with the dedicated ECG module of the
BITalino.

\subsubsection{EDA}\label{eda}

We measured arousal -- which relates to the intensity of an emotion and
varies from calm to excited (e.g. \emph{satisfied} vs \emph{happy}) --
using EDA. When measuring EDA, most accurate readings can be obtained
from the tip of the fingers. However, since it is difficult to
manipulate a tangible interface and controls while having hardware
attached to one's fingers, we acquire the signal from the palm of a
single hand instead. We assess skin conductance from two small
conductive thread patches sewn inside a fingerless glove (Figure
\ref{fig:eda}). Because the BITalino EDA amplifier was not sensitive
enough for signals acquired from the palm we made our own, replicating
the schematics described in \citep{Poh2010a}.

\subsubsection{Breathing}\label{breathing}

For breathing, we built a belt based on a stretch sensor (Figure
\ref{fig:breathing}). A conductive rubber band was mounted as a voltage
divider and connected to an instrumentation amplifier (Texas Instruments
INA128). As opposed to piezoelectric components, that are sensitive to
momentous speed instead of position, stretch sensors can directly map
users' chest inflation onto their avatar.

\subsubsection{EEG and Eye Blinks (EOG)}\label{eeg-and-eye-blinks-eog}

We built our own EEG helmet based on the open hardware OpenBCI board
(Figure \ref{fig:openbci}). To shorten setup time we used dry electrodes
-- the same TDE-201 as for ECG for the forehead, and elsewhere TDE-200
electrodes, which possess small protuberance that could go through the
hair. Using a stretchable headband, we restrained electrodes' locations
to the rim of the scalp to avoid difficulties with long-haired people.
In the 10-20 system, electrodes were positioned at O1, P7, F7, FP1, F8,
T8, P8 and O2 locations -- reference at T7, ground at FP2. We used
OpenViBE to analyze physiological data in real time\footnote{\url{http://openvibe.inria.fr/}}.
EEG signals were re-referenced using a common average reference.
Specific frequencies (see below) were extracted with a band-pass filter,
taking the log of the power of signals in order to normalize indices.
Eye blinks were detected when the signal, after DC drift removal,
exceeded 4 times the variance in the F8 electrode. We used the following
metrics:

\begin{itemize}
\itemsep1pt\parskip0pt\parsep0pt
\item
  Vigilance: appoints for the ability to maintain attention over time.
  We use the ratio between beta frequency band (15-20Hz) and theta + low
  alpha frequency band (4-10Hz) for all electrodes \citep{Oken2006}.
\item
  Workload: increases with the amount of mental effort required to
  complete a task. We use the ratio between delta + theta band (1-8Hz)
  in near frontal cortex (F7, FP1, F8, T8) and wide alpha band (8-14Hz)
  in parietal + occipital cortex (P8, P7, O2, O1)
  \citep{Antonenko2010, Schober1995}.
\item
  Meditation: we used instantaneous phase locking value between front
  (FP1, F7, F8) and rear (O1, P7, P8) parts of the brain in alpha + beta
  bands (7-28Hz) \citep{Frey2014b} -- mindfulness and body focus
  practices decrease the synchronization while transcendental practice
  increases it.
\item
  Valence: designates the hedonic tone of an emotion and varies from
  positive to negative (e.g. \emph{frustrated} vs \emph{pleasant}). We
  use the ratio between the EEG signal power in the left (F7, P7, O1)
  and right (F8, P8, O2) cortex in the alpha band (8-12Hz)
  \citep{Molina2009}.
\end{itemize}

In earlier iterations of the system we tested the use of an Emotive Epoc
headset to account for brain activity. The Epoc is a consumers-oriented
EEG device, easier to install than medical headsets that use gel.
However, it still requires a saline solution that tends to dry over
time, causing additional installation time between users. Moreover, good
signal quality was next to impossible to obtain with long haired
persons. Another downside of consumers-oriented EEG headsets is that
they usually conceal signal processing behind proprietary algorithms,
with little scientific evidence on what is truly measured. While
building a tailored EEG helmet, we took the upper hand on the whole
pipeline. With access to raw EEG signals, we looked into the literature
to match the inner state we wanted to measure with actual neurological
markers.

\subsection{Feedback}\label{feedback}

The feedback consists in both a support and a shape.

\subsubsection{Support}\label{support}

3D printing a tangible avatar is a powerful incentive for customization.
While the version of the system that we deployed in the scientific
museum used an already modeled and 3D printed incarnation of Tobe
because of time constraints, a user of the system could change the
parametric model in order to obtain an avatar that pleases her. The
process would be similar to how the appearance of a Nintendo ``Mii'' can
be tuned, except for the tangibility. As a tradeoff between preparation
time and customization, we prototyped a ``Mr.~Potato Head'' version of
Tobe, with parts ready to be assembled (Figure \ref{fig:potato}).

\subsubsection{Shape}\label{shape}

\begin{figure*}
  \centering
  \subfloat[]{\includegraphics[height=3.3cm]{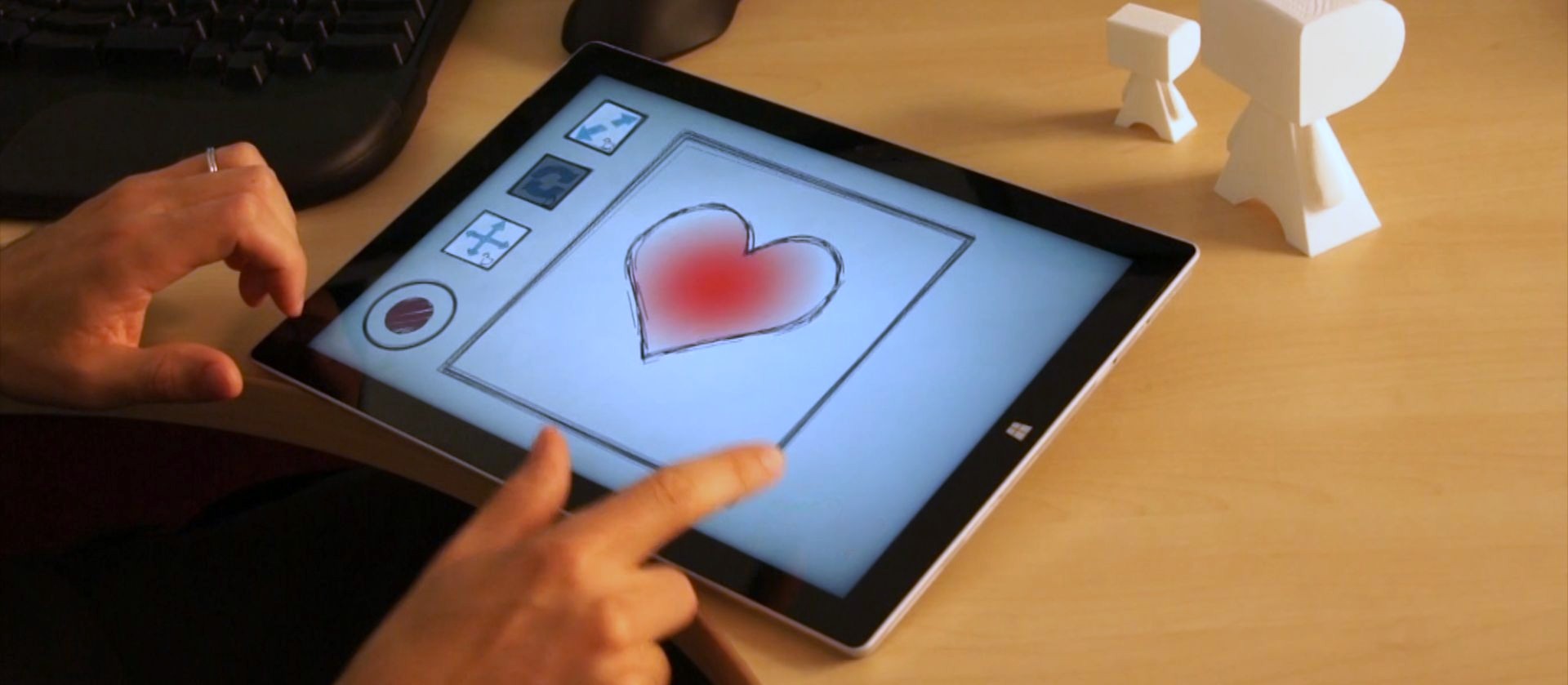}\label{fig:animator}}\hfill
  \subfloat[]{\includegraphics[height=3.3cm]{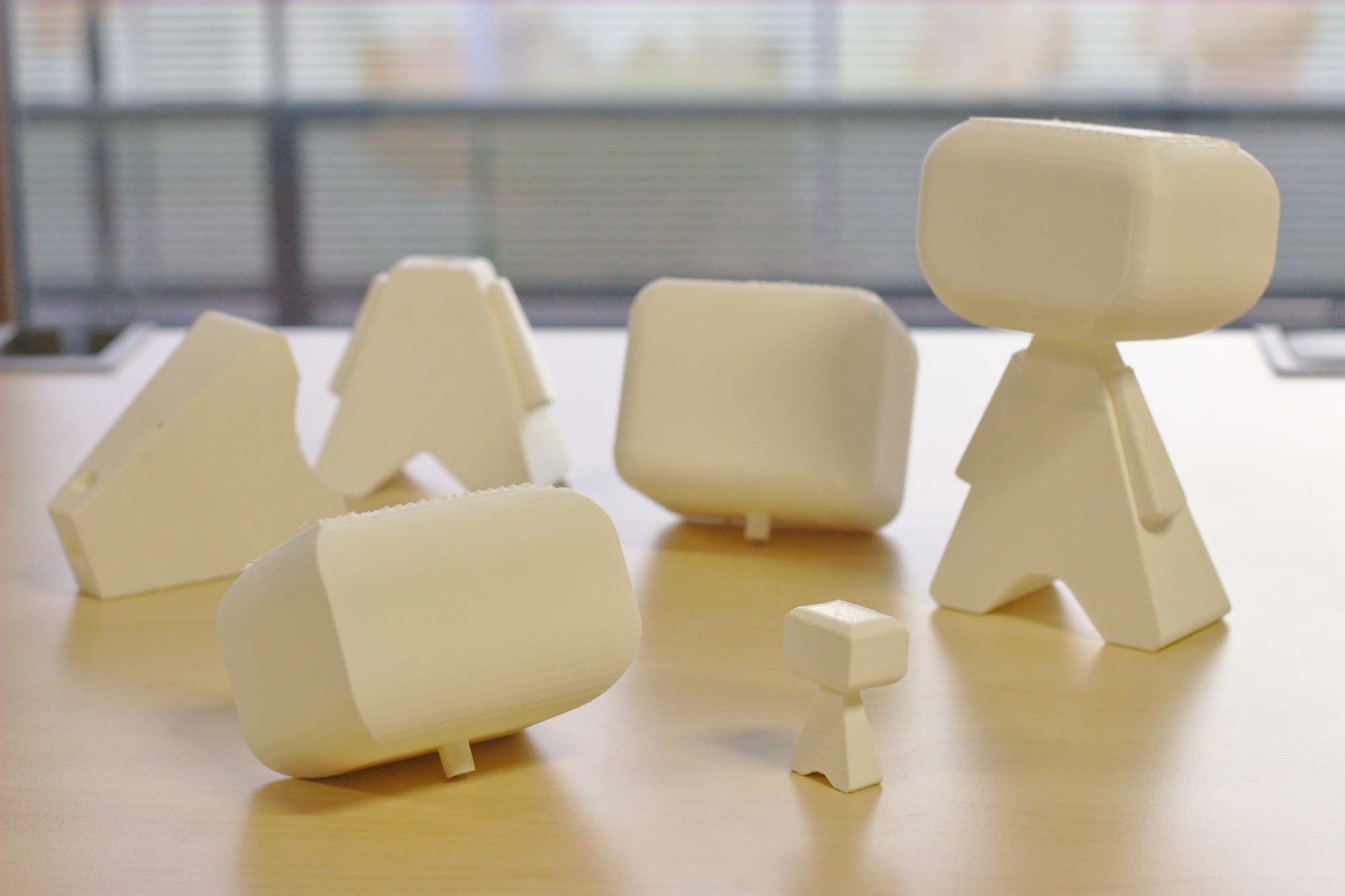}\label{fig:potato}}\hfill
  \subfloat[]{\includegraphics[height=3.3cm]{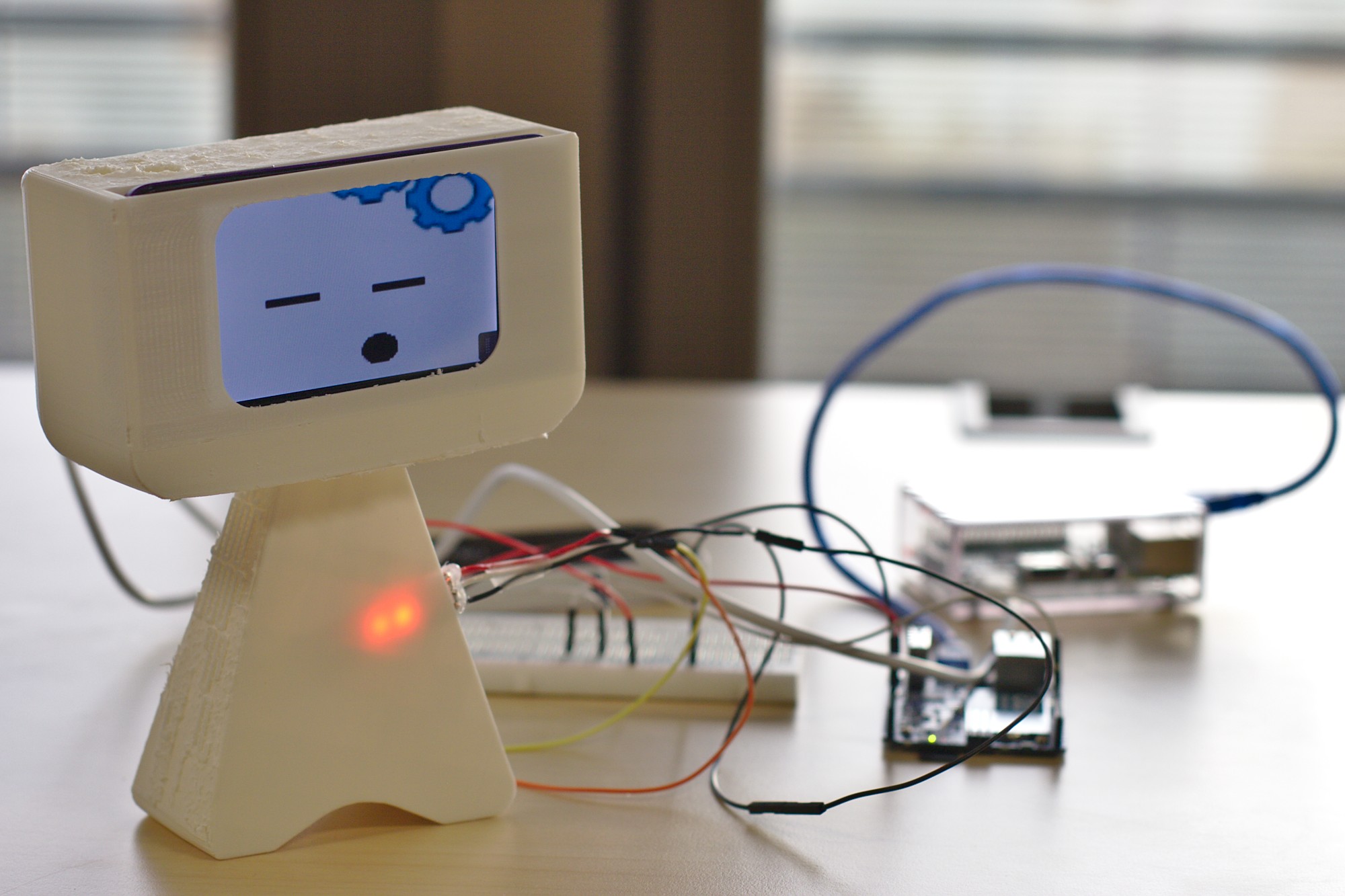}\label{fig:droid}}\hfill
  \caption{\emph{a}: Simple multitouch animator allowing users to create and animate visual feedback. \emph{b}: Customizing the tangible support of Tobe can be achieved using modular body pieces; \emph{c}: it is also possible to embedded electronics inside the support to have a standalone Tobe.}\label{fig:customization}
\end{figure*}

The visualization of users' signals are displayed onto Tobe using
Spatial Augmented Reality (SAR), as introduced by Raskar et al.
\citep{Raskar2001}. SAR adds dynamic graphics to real-world surfaces
using projected light. Despite external hardware -- i.e.~a projector and
eventually a tracking device (Figure \ref{fig:capsciences}) -- SAR is an
easy solution to prototype a system, faster to deploy than putting
actual screens in users' surroundings. The augmentation occurred within
vvvv\footnote{\url{http://vvvv.org/}}, a software that uses real-time
visual programming to render 3D scenes. We used a LG PF80G projector of
resolution 1920x1080 and the tracking of Tobe was achieved with an
OptiTrack V120:Trio, running a 120 FPS with an overall latency of 8.3ms
and a precision of 0.8mm. The projector was calibrated with the
OptiTrack using OpenCV's camera calibration function.

As an alternative to SAR, Tobe can be embedded with small screens, LEDs,
actuators and small electronics components so that it represents a
standalone unit. We already have a proof of concept of such an
implementation thanks to the easiness and accessibility of the building
blocks that go with the Arduino platform and the Raspberry Pi (Figure
\ref{fig:droid}).

\subsubsection{Customization}\label{customization}

We conceived a GUI that let users draw a picture and animate it
according to their wishes. The animator is touch based; users press a
``record'' button and animate the picture with gestures (Figure
\ref{fig:animator}). Once done, the animation's timeline is
automatically mapped to the chosen signal. This animator is kept simple
on purpose, it is designed for novice users and as such must remain easy
to understand and operate for someone not familiar with animation. Only
three basic operations are currently supported -- scaling, rotation and
translation -- and yet it is sufficient to generate meaningful
animations. For example, scaling makes a heart beat, translation moves a
cloud along respiration and rotation spins cogs faster as workload
increases. An advanced tool such as Photoshop has already been
integrated as a proof of concept, but the simplicity of the current GUI
does not impede users' creativity and is already sufficient to enable a
tailored feedback.

\section{Tobe in the Wild}\label{tobe-in-the-wild}

We used and tested Tobe in two different applications: as a
demonstration in a scientific museum and as a multi-user biofeedback
device for relaxation and empathy.

\subsection{Tobe in a public
exhibition}\label{tobe-in-a-public-exhibition}

Using a co-design approach, we intervened in a scientific museum over
two half days, proposing to passersby to try out Tobe. Around twelve
persons tested the system. We built the sensors and prepared the signal
processing beforehand because these steps require hardware and
expertise. Five high-level metrics were selected: workload, vigilance,
meditation, valence and arousal. These metrics were chosen because the
general public showed interest in them (meditation and emotions) or
because they could benefit from being better known (workload and
vigilance). Due to the short duration of our exhibitions, we also set
the corresponding feedback (both support and shape), according to the
outcome of the questionnaires about people's representations.

\begin{figure}
\centering
\includegraphics[]{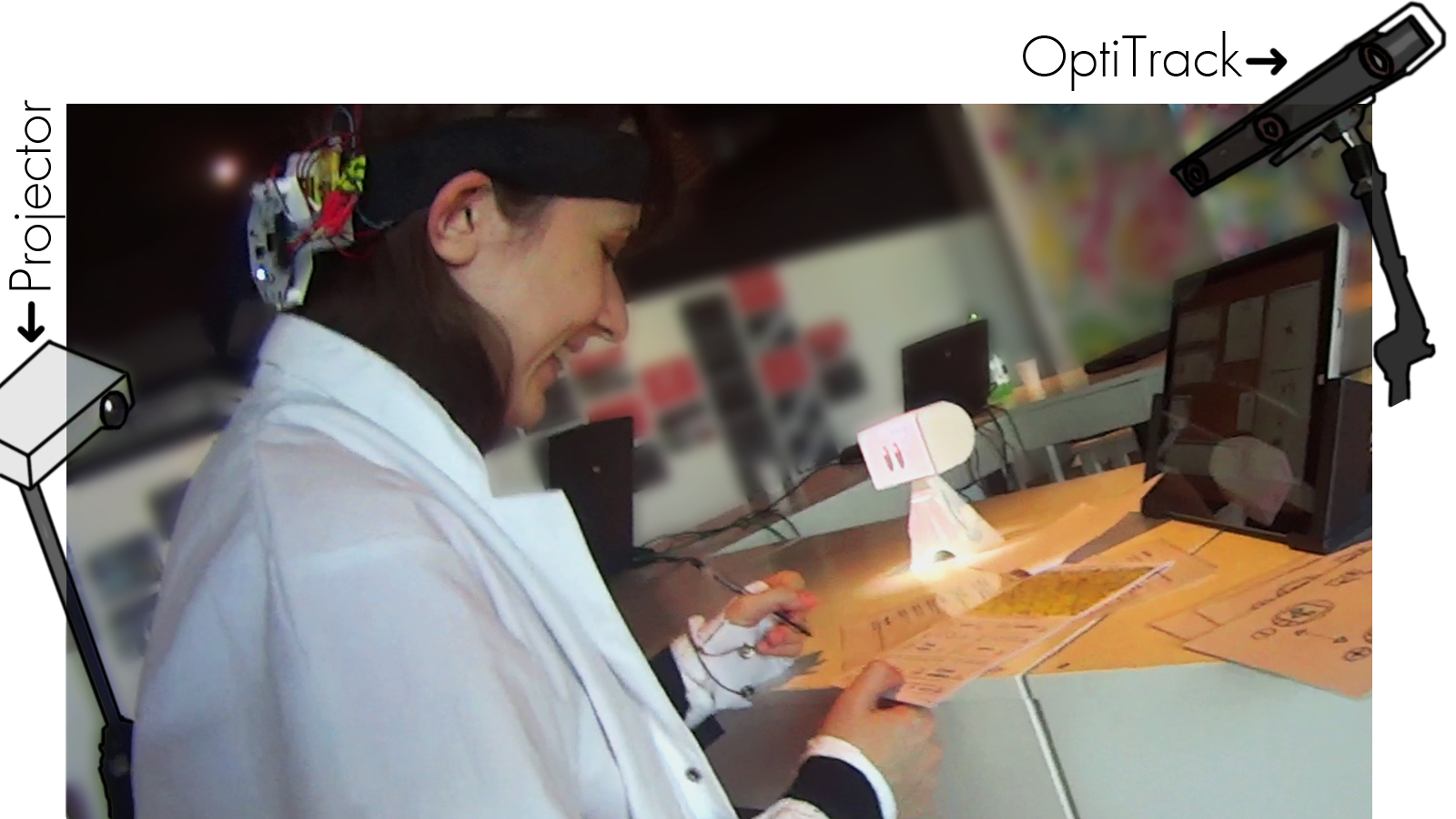}
\caption*{In a scientific museum, various activities were proposed to
visitors in order to prompt self-investigation. The setup consists of a
projector handling the augmentation and an OptiTrack for the
tracking.}\label{fig:capsciences}
\end{figure}

After we equipped participants, we gave them ``activity cards'', a
collection of scenarios designed to modify their inner state and to
prompt self-investigation (Figure \ref{fig:capsciences}). In particular,
these card asked them to perform riddles and arithmetic problems
involving working memory to increase their workload \citep{Frey2014a};
to look at cute and \emph{less} cute images to change their valence and
arousal levels (a typical emotion elicitation approach
\citep{Molina2009}), to perform a breathing exercise to help them
meditate, and to play a ``Where's Waldo?'' game to increase their
vigilance \citep{Frey2014a}. These sole cards sufficed to engage
participants for a few tens minutes without our intervention.

We created the activity cards after our first intervention in the
museum. There were some candies left at disposal next to Tobe to lure
museum's visitors to our booth. At some point, one user wanted to see
how different tastes affected the emotional valence that was displayed
on Tobe. This proved to be a fun activity for him -- and for the people
around. Having such goal in mind was an effective way to drive
participants.

One degree of freedom was left to users by the mean of a graphical
interface (see Figure \ref{fig:trailer}). They had to manipulate the GUI
on a nearby tablet to drag and drop visualizations on predefined anchor
points. Users could customize some of Tobe's aspects (eyes and heart
rate feedback) and among the 5 high-level metrics available, they
selected which one to study at a particular time. When at first we
tested Tobe with \emph{no} degree of freedom -- i.e.~all metrics were
displayed altogether -- we realized that users were too passive and
quickly overwhelmed. The GUI helped to focus and engage users.

To further engage users, Tobe was tracked and participants were asked to
put Tobe on a spotlight to ``awake'' it -- i.e.~to start physiological
signals' streams. The action of bringing life to an inanimate puppet
goes well with making the world ``magical'' again \citep{Rose2014}, that
is to say to use the power of abstraction of modern computer science in
order to bring back awe. The aim is not to take benefit of ignorance but
to strengthen the amazement that technology can offer. We were ourselves
pleasantly disturbed and surprised when we happened to hold in our hands
a representation of our beating heart during some routine test. Suddenly
the relationship with the digital content felt different, truly
tangible.

\subsection{Tobe for multi-users
relaxation}\label{tobe-for-multi-users-relaxation}

\begin{figure}
\centering
\includegraphics[]{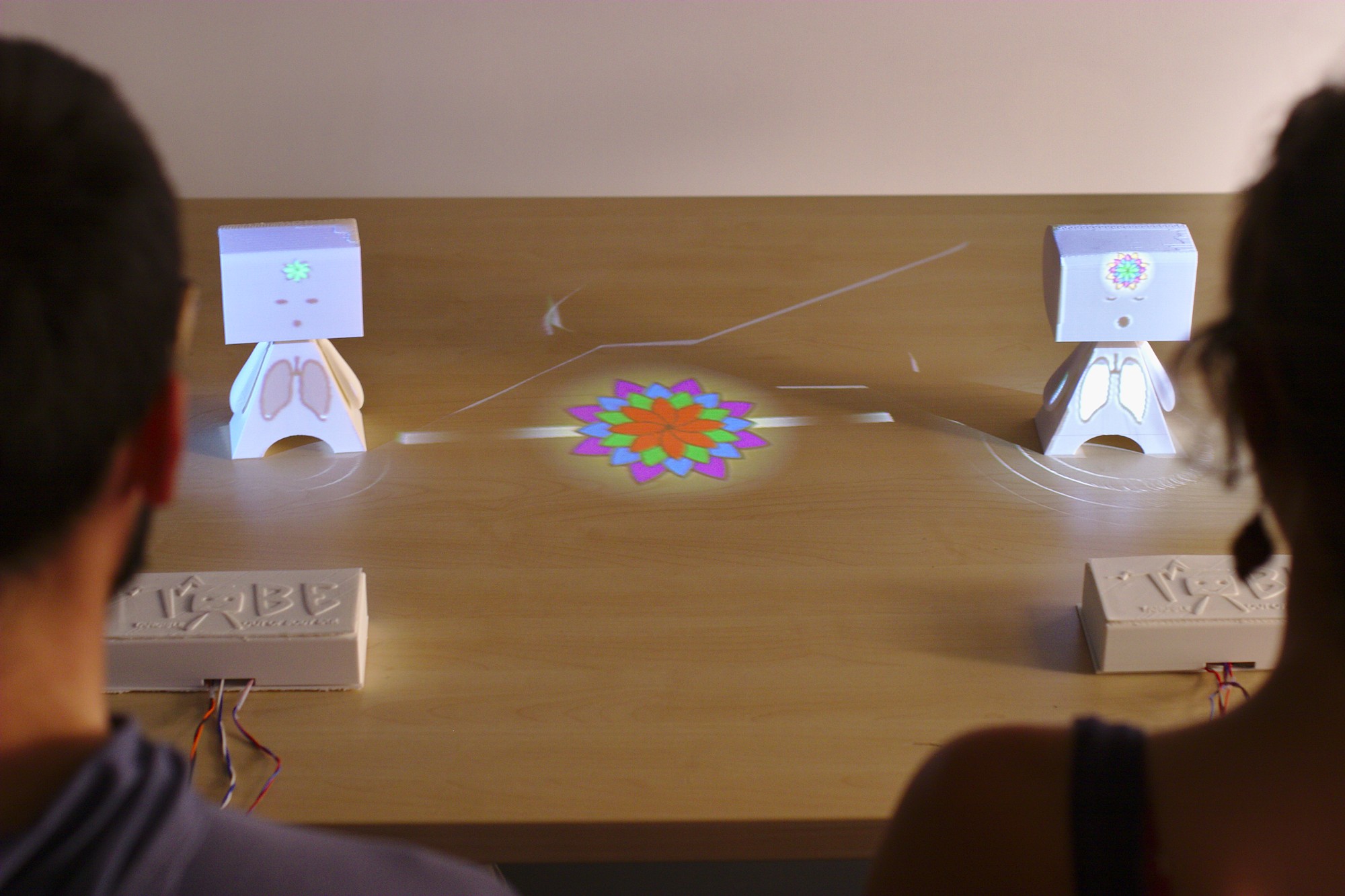}
\caption*{Multi-users application: relaxation through cardiac
coherence.}\label{fig:coherence}
\end{figure}

We tested Tobe as a relaxation device for two users (Figure
\ref{fig:coherence}). The objective was to see if Tobe could be used
both as a biofeedback tool and for collaboration.

\subsubsection{Implementation}\label{implementation}

This version of Tobe relies only on respiration and heart rate
variability. It relates to cardiac coherence: when someone takes deep
breaths, slowly ($\approx$ 10s periods) and regularly, her or his heart
rate (HR) varies accordingly and the resulting state has positive impact
on well-being \citep{McCraty2009}. During cardiac coherence, HR
increases slightly when one inhales and decreases as much when one
exhales. We took the magnitude squared coherence between HR and
breathing signals over 10s time windows as a ``relaxation'' index.

Sensors consisted in a breathing belt and in a pair of elastic bands
around the wrists to measure ECG. We used OpenBCI instead of BITalino to
measure ECG and breathing in order to get more accurate readings.
Indeed, the OpenBCI amplifier has a resolution of 24 bits instead of 10
for the BITalino.

There were two Tobes on the table, one for each participant. They were
not tracked. Breathing activity was pictured with inflating lungs onto
the torso; cardiac coherence with a blooming flower onto the forehead.
The synchronicity \emph{between} subjects -- users' heart rates varying
at the same pace -- was represented with a similar but bigger flower
projected between both Tobes. Additionally, ``ripples'' on the table,
around Tobes' feet, matched heart beats.

\subsubsection{Protocol}\label{protocol}

We asked 14 participants, by pairs, to come and use Tobe to reach
cardiac coherence -- 6 females, 8 males, mean age 25.3 (SD=2.8).
Participants were coworkers from the same research institute and already
knew each other. Participants were seated on each side of a screen and
instructed not to talk to each other. We presented them the cardiac
coherence activity as a relaxation exercise. Afterwards, we equipped
them and turned the system on.

The experiment comprised of three sessions of 5 minutes. During the
first session, participants had to individually learn how to reach
cardiac coherence. A smaller screen on the table prevented them to see
each other's Tobe. They had to imitate the breathing pattern given by a
gauge going up and down in 5s cycles onto Tobe's body. The lights of the
room were dimmed to facilitate a relaxation state and each participant
was given headphones playing back rain sounds.

After the training session, the screen separating the two Tobes was
removed. Participants were then instructed to repeat the same exercise
as before, but without the help of the gauge. They could see their
colleague's Tobe. However, there was no interaction between them at this
stage -- it served as a transition between a self-centered task and a
collaboration task.

During the third session, participants were instructed to synchronize
their \emph{hearts}. In order to do so, they had to both reach cardiac
coherence while breathing on the same rhythm -- with no other way to
communicate than using their Tobes.

After this final session, we gave questionnaires to participants and
conducted informal interviews with them to gather feedback about their
experience.

\subsubsection{Results}\label{results}

From the questionnaires, that took the form of 5-points Likert scales,
participants reported that they were more relaxed after the end of the
session: 4.36 on a scale ranging from 1 ``much \emph{less} relaxed'' to
5 ``much \emph{more} relaxed'' (SD=0.74). Beside the fact that Tobe
acted as an effective biofeedback device, the experiment was also a
chance to introduce participants to activities centered around
well-being, as few of them were practicing relaxation or meditation in
their daily life -- 1.93 score (SD=1.44) with 1 ``never'' and 5
``regularly''.

During the interviews, the participants reported that they appreciated
the feedback, saying that it formed a coherent experience --
e.g.~ripples on the table and sounds of rain. Among the few that were
practicing yoga regularly, one praised how Tobe favors learning-by-doing
over wordy and disrupting instructions but had troubles to follow the
10s breathing cycle since it differed from his usual practice. We had
mixed reviews about the visualization associated to breathing, mostly
due to the mapping between Tobe's lungs and the measured thoracic
circumference being dynamically adapted over time rather than calibrated
per user with a min/max. Because of that, some users had to draw their
attention away from the breathing patterns in order to achieve cardiac
coherence. These two last issues could be resolved by giving users
access to the signal processing through our toolkit.

We received comments about how a qualitative and ambient feedback
(blooming flower) fostered a better focus on the activity compared to
the use of quantitative metrics which are an incentive for competition.
Indeed, apart from some comparisons made during the second session,
participants did use their Tobes for collaboration. Users described how
they use the respiration of their partner to get in sync during the
third stage -- usually by waiting before inhaling. One participant
described how she tried to ``help'' her companion when he struggled to
follow. Another retold how she quickly resumed her regular breathing
when she saw that a brief hold troubled her colleague. More playful, a
participant laughed afterward at how he purposely ``tricked'' twice his
partner. Even with a communication channel as basic as the display of
thoracic circumference, rich interactions emerged between participants
over a short period -- 5 minutes that felt like less for many of them.

Overall these findings suggest that Tobe could be employed as a proxy
for interpersonal communications and that it has an interesting
potential for enhancing well-being.

\section{Discussion and potential
applications}\label{discussion-and-potential-applications}

The toolkit that we propose uses physiological sensors to let people
grasp and share their inner states. Even if the modules we chose promote
the inclusion of novices -- e.g.~visual programming that could be easily
extended in OpenViBE or vvvv, they can be switched to other components
that would suit more experienced users -- e.g.~Matlab for signal
processing. Moreover, we explored multiple ways to support the feedback
via SAR and embedded electronics. We think that SAR provides a very
flexible and high-resolution option for prototyping different feedbacks.
However, for long term usages and deployment, we would recommend the use
of more embedded solutions since SAR still suffers from a more
complicated installation and occlusion issues.

Beside the applications that we tested in the wild, we drew usages for
Tobe by exploring different design dimensions. We considered on the one
hand the number of users, Tobes and external observers involved, and, on
the other hand the time and space separating the feedback and the
recordings. The following scenarios could be used by researchers and
enthusiasts to explore the possibilities offered by our toolkit.

\subsubsection{One User}\label{one-user}

Tobe can be used as a biofeedback device with a specific goal --
e.g.~reduce stress -- or to gain knowledge about one self. A feedback
about workload and vigilance would prevent overwork. Insights gathered
from an introspection session with Tobe could also be employed to
\emph{act} better. For example, it might be useful to realize that you
are irritated before answering harshly to beloved ones.

\subsubsection{One User and Observer(s)}\label{one-user-and-observers}

Tobe could be used in a medical context. Indeed, in stroke
rehabilitation, patients with motor disabilities may regain mobility
after long and difficult sessions of reeducation. However, occasional
drawbacks may create anxiety and a counterproductive attitude towards
therapy, which leads to even more anxiety. A Tobe could help patients
and therapists acknowledge this affective state and break this vicious
circle. Autistic persons could also benefit from using Tobe since it is
difficult for them and their relatives to gauge their inner state.
Explicit arousal could help their integration into society. An offline
experiment -- i.e.~after signals were recorded -- pointed to this
direction \citep{Hedman2012}.

\subsubsection{Multiple Users and Tobes}\label{multiple-users-and-tobes}

Using Tobe as an alternate communication channel during casual
interactions would help to explore connections with relatives, discover
and learn from strangers or improve collaboration and efficiency with
coworkers. This has been partially explored through the ``Reflect
Table'', which gives a feedback about the affective state of meeting
participants \citep{Bachour2010}; and a bicycle helmet that displays the
heart rate of the wearer to the other cyclists nearby has been proposed
to support social interactions during physical efforts
\citep{Walmink2014}.

\subsubsection{Archetype of a Group}\label{archetype-of-a-group}

Tobe could summarize the state of a group. A real-time feedback from the
audience would be a valuable tool for every speaker or performer. To
pace a course, a teacher could use one Tobe as an overall index that
aggregates the attention level of every student in the classroom.
Through behavioral measures and with a feedback given afterward, this
was investigated in \citep{Raca2013}.

\subsubsection{Time and Space}\label{time-and-space}

One could want to analyze or to recall inner states after an event. Tobe
could replay how one actually felt alongside a video of a cherished
moment. If it is not time but space that separates a Tobe from its
owner, imagine a distant relationship where the Tobe on your desk slowly
awakens as the sun rises in the timezone of your beloved one -- and you
would wait for Tobe's vigilance to increase to a sufficient level before
you pick up your phone for a chat, knowing that your soul mate is a bit
grumpy at the beginning of the day. Besides this theoretical view, it
has been hinted that even low-level physiological signals could enhance
telepresence \citep{Lee2014}.

\section{Conclusion}\label{conclusion}

We have presented an open system aimed at externalizing physiological
signals and mental states in order to offer users a shared ``out-of-body
experience''. This system covers the entire pipeline, from signals'
acquisition to their visualization. The open nature of the toolkit may
be used to introduce STEM disciplines (Science, Technologies,
Engineering and Mathematics) to the general public through inquiry-based
learning, while end usages can steer them to cognitive science,
psychology and humanities. Future work will include testing Tobe in
classrooms or public workshops where users will be invited to build
their own self-representation from the ground up, including the tangible
support, sensors and the feedback design. Longer usages of the toolkit,
over multiple days or weeks, will also be the opportunity to strengthen
signal processing in order to provide more reliable mental states that
could be displayed between users. Tobe could be used to ease social
interaction or to foster empathy towards others. Giving users the tools
to investigate their own bodies and mind is a good way to empower them
and prompt self-reflection.

\section{Acknowledgments}\label{acknowledgments}

We thank Christelle Godin, Matthew S. Goodwin, Pierre-Alain Joseph, Éric
Sorita, Dominique Dionne, Oussama Azibou and Cap Sciences for their
ideas and support during this project.

\balance{}

\bibliography{tobe}

\end{document}